\documentclass[letterpaper]{article} 
\usepackage{aaai2026}  
\usepackage{times}  
\usepackage{helvet}  
\usepackage{courier}  
\usepackage[hyphens]{url}  
\usepackage{graphicx} 
\usepackage{natbib}  

\usepackage{makecell}

\usepackage{amsmath} 
\usepackage{amssymb}
\usepackage{booktabs} 
\usepackage{multirow} 
\usepackage{tabularx} 
\usepackage{ragged2e} 

\usepackage{xcolor}
\usepackage{caption} 
\usepackage{algorithm}
\usepackage{algorithmic}

\usepackage{colortbl} 
\usepackage{newfloat}
\usepackage{listings}
\DeclareCaptionStyle{ruled}{labelfont=normalfont,labelsep=colon,strut=off} 
\floatstyle{ruled}
\newfloat{listing}{tb}{lst}{}
\floatname{listing}{Listing}
\usepackage{hyperref} 

\title{AHAMask: Reliable Task Specification for Large Audio Language Models\\ Without Instructions}
\author{
    Yiwei Guo\textsuperscript{\rm 1}, Bohan Li\textsuperscript{\rm 1}, Hankun Wang\textsuperscript{\rm 1}, Zhihan Li\textsuperscript{\rm 1}, Shuai Wang\textsuperscript{\rm 2}, Xie Chen\textsuperscript{\rm 1}, Kai Yu\textsuperscript{\rm 1}\thanks{Corresponding author.\\Code link: {https://github.com/X-LANCE/SALMONN-AHAMask}}
}
\affiliations{
    \textsuperscript{\rm 1}X-LANCE Lab, School of Computer Science, Shanghai Jiao Tong University, China \\
    MoE Key Lab of Artificial Intelligence, Jiangsu Key Lab of Language Computing, China\\
    \textsuperscript{\rm 2}School of Intelligence Science and Technology, Nanjing University, China

    \{yiwei.guo, kai.yu\}@sjtu.edu.cn
}

\newcommand{\MethodName}{AHAMask}

\begin{document}

\maketitle

\begin{abstract}
Although current large audio language models (LALMs)  extend text large language models (LLMs) with generic acoustic understanding abilities, they usually suffer from prompt sensitivity, where different instructions of the same intention can yield drastically different outcomes. In this work, we propose \MethodName{}, where we simply mask some of the attention heads in the decoder-only LLM backbone of LALMs, to trigger specific acoustic task functionalities without instructions. These masks are efficiently obtained by training on an LALM, with the number of trainable parameters equal to the attention head count in its LLM backbone. We show by experiments that applying such selective attention head masks achieves comparable or even better performance than using instructions, either on single or composite tasks. Besides achieving reliable acoustic task specification for LALMs, this also reveals that LALMs exhibit certain ``functional pathways'' in their attention heads. 
\end{abstract}


\section{Introduction}
\begin{figure*}
    \centering
    \includegraphics[width=0.99\linewidth]{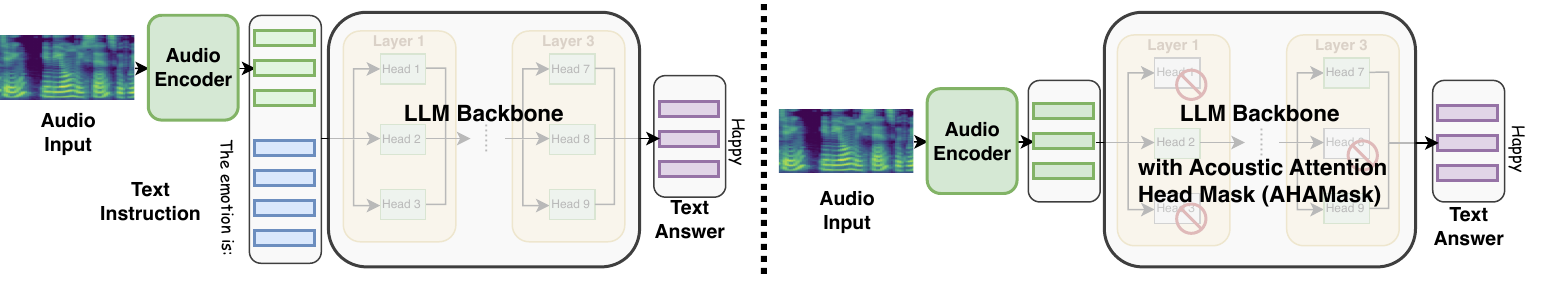}
    \vspace{-0.05in}
    \caption{Diagram of a typical large audio language model. \textbf{Left}: the original model, which requires a text instruction to perform a specific task, but is sensitive to instructions. \textbf{Right}: the large audio language model with \MethodName{} (acoustic attention head mask), where only a set of attention heads are activated for a specific task. This frees the need for text instructions.}
    \label{fig:main}
    \vspace{-0.15in}
\end{figure*}

The rise of large language models (LLMs) has brought a revolution in the speech and audio processing domain, resulting in a mature paradigm of large audio language models (LALMs)~\cite{peng2024survey,su2025audiosurvey,yang2025towards,arora2025landscape}.
In this work, we consider LALMs which are Transformer-based~\cite{vaswani2017attention} models that can generate textual response to audio inputs and instructions.
They extend text-based LLMs to ``hear'' and further understand audio input, typically with an audio encoder as the acoustic sensory receptor, and an LLM as the ``brain'' to process and response according to the audio input.

A prominent advantage of these LALMs is their unified framework of processing all kinds of audio information.
Here, instructions play a key role in specifying the task of such LALMs.
With a unified surface, LALMs can give answers to different audio traits, e.g. spoken content, speaker gender, identity, emotion, or even complex audio and music events, only by changing the instructions.

However, the flexibility of natural language also brings the risk of \textbf{prompt sensitivity}.
Even targeted at a specific task, LALMs can still exhibit a large degree of sensitivity to provided instructions, even when these instructions have identical task requirements but only differ in linguistic forms or template specifications.
This sensitivity poses a significant challenge for LALMs, impacting both their ability to {accurately follow instructions and maintain consistent task performance}~\cite{peng2024survey}.
This problem is also a consensus in the broader LLM research field.
However, for LALMs, limited effort has been devoted to benchmarking prompt sensitivity, let alone addressing it.

Since instructions are expected to control the functionality of LALMs, we choose to directly control the functionality inside the Transformer~\cite{vaswani2017attention} structure of LALMs instead.
This relates to the general explainability and functional partitioning of decoder-only Transformers, which have been a trending research topic for text LLMs.
Recently, \citet{han2025heads} finds that simply masking certain attention heads in LLMs leads to specific task functionalities without instructions.
This selective attention head masking mechanism does not apply any modification to the model parameters, revealing the fundamental existence of ``functional pathway'' among the attention heads.
However, this discovery is limited only to narrow applications in the text modality instead of multi-modal LLMs like LALMs.
The acoustic functionalities in LALMs are fundamentally distinct from those of text LLMs, requiring a focus on multimodal alignment between continuous signals and discrete textual spaces, together with the comprehension of paralinguistic and non-linguistic information.
Therefore, it is worth investigating whether similar properties exist in LALMs.

Inspired by the previous works in text-based LLMs, we apply the method in \citet{han2025heads} and verify in this study that the ``acoustic functional pathways'' also exist in LALMs' attention heads.
Thus, we propose \MethodName{} (Acoustic Attention Head Mask) to achieve reliable task specification for LALMs, simply by masking some attention heads.
Fig. \ref{fig:main} depicts the difference between a typical LALM with instructions and with only \MethodName{}.
Importantly, the parameter count of our method is only equal to the number of attention heads in LALM's decoder-only LLM backbone, which is even \textbf{magnitudes smaller then previous parameter-efficient fine-tuning methods} (1-2k parameters vs. millions of parameters).
This means training an \MethodName{} is highly efficient.
As these masks are binary in inference stage, the storage overhead of such masks is also neglectable, e.g. 200 bytes for SALMONN~\cite{tang2024salmonn}.
\MethodName{} distinguishes itself from existing fine-tuning approaches by \textbf{reducing} the effective parameter count during inference, rather than maintaining or increasing it.
We conduct experiments across famous LALMs and classic audio and speech understanding tasks, including single and composite multi-hop tasks.
Our findings suggest that:
\begin{itemize}
    \item 
    Simply masking some of the attention heads in the LLM backbone frees the need of instructions to specify a task in LALMs.
    These masks are intrinsic in the LALMs, not depending on the inputs.
    \item On most single auditory tasks, LALMs with \MethodName{} perform comparably or even better than LALMs with natural language instructions.
    \item On composite tasks where LALMs typically struggle, \MethodName{} can guide the model to adhere to task requirements in a much more effective way than natural language instructions.
    \item The masks across different tasks exhibit varying degrees of correlation; for instance, more similar tasks tend to involve a greater overlap in activated attention heads.
    \item Acoustic functionalities are formed gradually if we activate attention heads in order based on their importance weights.
    In other words, the acoustic functional pathways are constructed collectively by the heads.
\end{itemize}

\begin{table*}[]
\centering
\small{
\begin{tabular}{@{}llll@{}}
\toprule
\multicolumn{1}{c}{\textbf{Task}} & \multicolumn{1}{c}{\textbf{Metric}} & \multicolumn{1}{c}{\textbf{Training Data}} & \multicolumn{1}{c}{\textbf{Test Data}} \\ \midrule
\multicolumn{4}{l}{\textit{\textbf{Single tasks}}}\\
ASR \scriptsize{(Automatic Speech Recognition)} & WER (Word Error Rate) & LibriSpeech & LibriSpeech test-clean \& test-other \\
GR \scriptsize{(Gender Recognition)} & ACC (Accuracy) & LibriSpeech train-clean-100 & LibriSpeech test-clean \\
SER \scriptsize{(Speech Emotion Recognition)} & ACC (Accuracy) & IEMOCAP Session1-4 & IEMOCAP Session 5 \\
ASV \scriptsize{(Automatic Speaker Verification)} & ACC (Accuracy) & VoxCeleb1 & VoxCeleb1-O(cleaned) \\
AAC \scriptsize{(Automatic Audio Captioning)} & METEOR \& ROUGE-L & AudioCaps & AudioCaps test \\
S2TT \scriptsize{(Speech to Text Translation)} & BLEU-4 & CoVoST2 (en$\rightarrow$zh) & CoVoST2 (en$\rightarrow$zh) test \\
OSR \scriptsize{(Overlapped Speech Recognition)} & WER (Word Error Rate) & \makecell[l]{Libri2Mix mix-clean \\train-100 \& train-360} & Libri2Mix (mix-clean) test \\ 
\midrule
\multicolumn{4}{l}{\textit{\textbf{Composite tasks}}}\\
\makecell[l]{GR\texttt{|}ASR, ASR\texttt{|}GR, \\ \{``ASR'': , ``GR'':\} }&  \makecell[l]{IFR (Instruction Following Rate),\\ WER, ACC} & LibriSpeech train-clean-100 & LibriSpeech test-clean \\
\bottomrule
\end{tabular}
}
\vspace{-0.05in}
\caption{Audio understanding tasks, metrics, training data, and test data considered for LALMs in this paper.}
\label{tab:tasks}
\vspace{-0.15in}
\end{table*}

Generally, \MethodName{} leverages the fundamental and intrinsic property of acoustic functional pathways in LALMs, and addresses the sensitivity of task specification in LALMs in a highly parameter-efficient way.

\section{Background}
\subsection{Large Audio Language Models}

Popular LALMs in the current literature usually make use of an audio encoder and a pretrained backbone text LLM (see Fig. \ref{fig:main} left).
The audio encoders are self-supervised or supervised learning models to provide high-dimensional contextual representations of input audio signals. 
For example, the encoder of Whisper~\cite{whisper} and BEATs~\cite{beats} are used and compressed together by Q-former~\cite{li2023blip} in SALMONN~\cite{tang2024salmonn} for speech and general audio input, respectively.
These extracted embeddings are fed to an LLM backbone, which is typically initialized from a pretrained text LLM and finetuned on audio data, such as by LoRA~\cite{hu2022lora}.
After training, the backbone LLM is able to understand and answer questions on acoustic information in various perspectives, such as spoken content, prosody, speaker traits, and even audio events.
Recently, rapid progress has been made in this direction~\cite{hu-etal-2024-wavllm,lu24c_interspeech_desta,ding2025kimi,geng2025osum,goel2025audio,sakshi2025mmau}, making LALMs a successful and necessary component in artificial general intelligence.

\subsection{LLM Instruction Following and Prompt Sensitivity}
The prompt sensitivity issue in LLMs usually comes in the form of instruction non-following.
For text-only LLMs, several benchmarks~\cite{zeng2024evaluating, zhou2023instruction, qin2024infobench, lou2024large, wen2024bench} for evaluating instruction-following capabilities have been established. 
Optimization methods for improving instruction-following ability include steering vectors~\cite{stolfo2025improving,he2025saif}, supervised finetuning or reinforcement learning with sophisticated data pipelines~\cite{xu2024wizardlm,dong2025selfplay,an2025ultraifadvancinginstructionfollowing,agrawal2025enhancing}.
Additionally, prompt sensitivity is widely observed~\cite{chatterjee-etal-2024-posix, sclar2024quantifying, worstprompt_nips2024, zhuo-etal-2024-prosa, razavi2025benchmarking}; for instance, minor modifications such as altering punctuation marks alone can degrade LLM performance significantly~\cite{sclar2024quantifying}.
In contrast, efforts to benchmark instruction-following ability and sensitivity in LALMs are ongoing but comparatively limited~\cite{lu2025speech,gao2025ifevalaudio}.

\subsection{Functional Partitioning of LLMs}
In text-based LLMs, researchers have demonstrated that some Transformer modules exhibit certain functionality.
For example, some Transformers layers can be swapped for cross lingual transfer~\cite{bandarkar2025layer}.
It is also found that the neurons in feed forward networks (FFNs) store {knowledge}, such as language~\cite{zeng-etal-2025-converging}, concepts~\cite{rai2024investigation}, and tasks~\cite{xiao2024configurable}.
Analysis on the attention heads also show that they encourage instruction following~\cite{wu-etal-2024-language}, transport a compact representation of tasks~\cite{todd2024function}, mitigate knowledge conflicts~\cite{jin-etal-2024-cutting}, etc.
Recently, \citet{han2025heads} finds that learning-based attention head selection can form specific task functional pathways in LLMs without instructions.
These findings indicate LLMs possibly have some internal modularity.
Note that this also has some similarity with studies in functional partitioning of the brain from neuroscience~\cite{bertolero2015modular,wig2017segregated}.
To the best of our knowldge, such partitioning and modularity have not been explored in LALMs so far.

\section{\MethodName{}: Acoustic Attention Head Mask}


\label{sec:training}

\begin{figure*}
    \centering
    \includegraphics[width=0.99\linewidth]{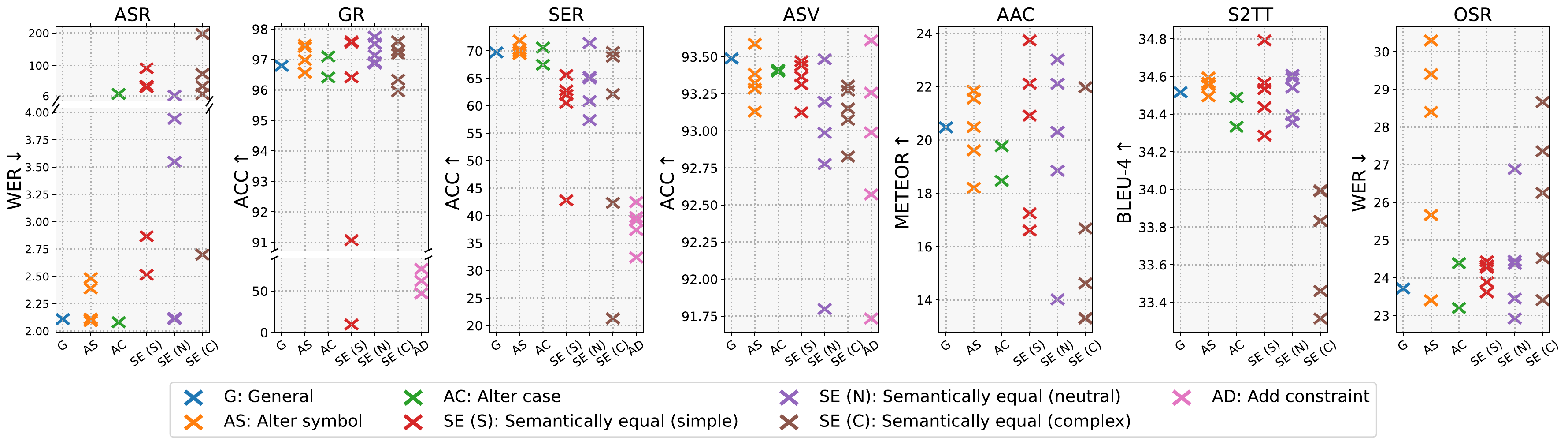}
    \vspace{-0.1in}
    \caption{Prompt sensitivity experiments on SALMONN. Different colors and columns denote different types of variations.}
    \label{fig:sensitivity}
    \vspace{-0.14in}
\end{figure*}

Since the backbone LLMs act as ``brains'' in LALMs, we follow  \citet{han2025heads} and seek for acoustic functional pathways inside those LLMs.
Modern LLMs are decoder-only Transformers, where each layer consists of multi-head attention (MHA), normalization, feed forward network and skip connections.
Denote $\mathbf X\in\mathbb R^{l\times d}$ as the input sequence with length $l$ and dimension $d$, the attention operation at the $i$-th layer and $j$-th attention head is defined as
\begin{equation}
    \mathbf Y^{(i,j)}=\operatorname{Softmax}\left(\frac{\mathbf X\mathbf W_Q^{(i,j)}\left(\mathbf X\mathbf W_K^{(i,j)}\right)^T}{\sqrt{d_{\text{head}}}}\right)\mathbf X\mathbf W_V^{(i,j)},
\end{equation}
where $\mathbf W_Q^{(i,j)},\mathbf W_K^{(i,j)},\mathbf W_V^{(i,j)}$ are query, key and value projections with shape $d\times d_{\text{head}}$, respectively.
The output of an MHA module is then the weighted sum of all attention head outputs in that layer, i.e. $\operatorname{MHA}_i(\mathbf X)=\sum_{j=1}^h \mathbf Y^{(i,j)}\mathbf W_O^{(i,j)}$ where $\mathbf W_O^{(i,j)}\in\mathbb R^{d_{\text{head}}\times d}$ is the output projection.
These MHA modules are the most important part of a Transformer, and have been a long-standing research topic in LLMs~\cite{vaswani2017attention,lin2022survey,zhao2023survey,zhao2024explainability,zheng2024attention}.

As a single attention head is the basic granularity in the MHA operation, we try to find out the set of attention heads that can trigger a specific functionality.
Let $\mathcal M\in\{0,1\}^{n\times h}$ be a set of attention head mask indicators, where $n,h$ are the number of layers in the decoder-only Transformer and the number of heads in each layer, respectively.
Denote $m_{i,j}\in \mathcal M$ as the binary indicator of the $i$-th layer and $j$-th attention head in that layer.
We now modify the MHA operation to
\begin{equation}
    \operatorname{MHA}_i(\mathbf X,\mathcal M)=\sum_{j=1}^h m_{i,j}\mathbf Y^{(i,j)}\mathbf W_O^{(i,j)}.
    \label{eq:masked-mha}
\end{equation}
In other words, the attention heads are activated when $ m_{i,j}\in \mathcal M$ is 1, or masked vice versa.
Note that due to the existence of skip connections, the computation graph won't be cut off even if all the $m_{i,j}$ in the $i$-th layer are 0.

With this notation, we can treat $\mathcal M$ as the trainable parameters, and use Eq.\eqref{eq:masked-mha} to train an \MethodName{}.
As $\mathcal M$ is a discrete variable, we use Gumbel-Sigmoid~\cite{jang2017categorical,geng-etal-2020-selective,han2025heads} for gradient estimation. Specifically, let $\mathbf M\in\mathbb R^{n\times h}$ be the mask logits of all heads. 
In the forward process, the discrete mask $\mathcal M$ is obtained by
\begin{equation}
    \mathbf S=\sigma\left(\frac{\mathbf M+\mathbf G}{\tau}\right) , ~~ \mathcal M=\mathbb I\left(\mathbf S\ge 0.5\right)
\end{equation}
where $\sigma(\cdot)$ is the sigmoid function, $\mathbb I(\cdot)$ is the indicator function, and $\mathbf G\in\mathbb R^{n\times h}$ is Gumbel noise sampled by $g_k=-\log(-\log \mathbf \epsilon_k) ,\forall g_k\in\mathbf G$ using $\epsilon_k\overset{\mathrm{i.i.d}}{\sim}\operatorname{Uniform}(0,1)$.
In the backward pass, gradients on $\mathcal M$ (the hard mask) is grafted to $\mathbf S$ (the soft mask), using straight-through estimator~\cite{bengio2013estimating}.
Scalar $\tau>0$ is a temperature hyperparameter used to control the sharpness of Gumbel approximation.
Therefore, we only train the mask logits $\mathbf M$ using gradient descent.
In inference, we obtain the discrete mask as $\mathcal M=\mathbb I(\mathbf M\ge 0)$.
Also, $\mathbf M$ assigns each attention head an importance weight, which can be utilized to explore the gradual formation of functional pathways.
Typically, with approximately 3 to 15 billion parameters,  LALMs contain only a few thousand attention heads within their LLM backbone.
This means the number of trainable parameters in \MethodName{} remains minimal, and the training process is highly efficient.

We train \MethodName{} on specific downstream tasks without instructions. Formally, for a given task $\mathcal T$, the training dataset can be represented as $\mathcal D_{\mathcal T}=\{...,(\texttt{Audio}_k,\texttt{Text}_k),...\}$, where $\texttt{Audio}_k$ is an audio clip input, and $\texttt{Text}_k$ represents the corresponding textual response for the task.
The training objective employs the standard cross entropy loss $\mathcal L_{\text{CE}}$ for next-token prediction applied to the tokens within $\texttt{Text}_k$.
Through this training paradigm, we anticipate that $\mathcal M$ will effectively identify the subset of attention heads associated with specific task functionalities within the LALM, without any instructions.

\section{Experiments}

\subsection{Experimental Setup}

\subsubsection{LALM Models}
In this work, we conduct studies on three well-known open-sourced LALMs: SALMONN~\cite{tang2024salmonn}, Qwen2Audio-Instruct, and Qwen2Audio-Base~\cite{chu2024qwen2audio}.
SALMONN incorporates Whisper and BEATs as acoustic encoders, and vicuna v1.1~\cite{vicuna2023} 13B as its backbone LLM.
Each of the two variants of Qwen2Audio use Whisper-style encoders and Qwen-7B~\cite{qwen} as their backbone LLM.
Qwen2Audio-Instruct is a supervised fine-tuned outcome of Qwen2Audio-Base for better instruction following ability. 
All the three models have exhibited good audio understanding abilities in various tasks as general LALMs~\cite{sakshi2025mmau}.

\subsubsection{Tasks and Datasets}
To discover atomic acoustic functionalities, we consider well-established benchmark tasks covering different aspects of audio understanding. 
These include automatic speech recognition (ASR) and gender recognition (GR) with LibriSpeech~\cite{librispeech}, speech emotion recognition (SER) with IEMOCAP~\cite{iemocap}, automatic speaker verification (ASV) with VoxCeleb1~\cite{voxceleb}, automatic audio captioning (AAC) with AudioCaps~\cite{kim2019audiocaps}, speech to text translation (S2TT) with CoVoST2~\cite{wang2020covost}, and overlapped speech recognition (OSR) with Libri2Mix~\cite{cosentino2020librimix}.
In GR, SER and ASV, LALMs are required to perform classification tasks , e.g. \textit{Male} or \textit{Female} in GR, and \textit{Yes} or \textit{No} in ASV.
In ASV, two speech utterances are concatenated with a 0.1s silence as a single input.
OSR requires LALMs to transcribe two overlapped speech utterances in any order.
In S2TT and AAC, LALMs should apply more complex abilities to understand the audio input, and output translations or captions.
These tasks span across purely linguistic (as ASR, OSR, S2TT), paralinguistic (SER, GR, ASV) to even general acoustic (AAC) understanding capabilities.

Moreover, we design composite tasks that require LALMs to perform multi-hop tasks in a specific format. 
In this study, we compose ASR and GR tasks and ask LALMs to perform them in order with separator symbol ``\texttt{|}'', or in json format.
We will show that LALMs usually fail on these tasks even with clear natural language instructions, while applying \MethodName{} can introduce significant performance boosts.
Table \ref{tab:tasks} lists all the tasks, metrics, and datasets used.

\subsubsection{Training Details}  
We initialize all the head logits in $\mathbf M$ by Gaussian $\mathcal N(4, 0.02)$ so that all heads are activated at first.
We apply an annealing schedule on Gumbel temperature $\tau$ so that it starts from 4.0, linearly decreases to 0.5 at 3k steps, and remains constant thereafter.
The learning rate is linearly warmed up from 1e-6 to 1e-2 within the first 3k steps, and decreased to a minimum of 1e-4 using cosine schedule afterwards.
All the original parameters from LALM are frozen, and $\mathbf M$ is the only trainable parameter.
All trainings are done on a single 65G Ascend 910B NPU device.

\begin{table*}[]
\centering
\small{
\begin{tabular}{@{}lcccccccc@{}}
\toprule
\textbf{Task} & \multicolumn{1}{c}{\textbf{ASR}} & \multicolumn{1}{c}{\textbf{GR}} & \multicolumn{1}{c}{\textbf{SER}} & \multicolumn{1}{c}{\textbf{ASV}} & \multicolumn{2}{c}{\textbf{AAC}} & \multicolumn{1}{c}{\textbf{S2TT}} & \multicolumn{1}{c}{\textbf{OSR}}\\ \cmidrule(r){1-1} \cmidrule(lr){6-7}
\textbf{Metric} & \multicolumn{1}{c}{\textbf{WER$\downarrow$}} & \multicolumn{1}{c}{\textbf{ACC$\uparrow$}} & \multicolumn{1}{c}{\textbf{ACC$\uparrow$}} & \multicolumn{1}{c}{\textbf{ACC$\uparrow$}} & \multicolumn{1}{c}{\textbf{METEOR$\uparrow$}} & \multicolumn{1}{c}{\textbf{ROUGE-L$\uparrow$}} & \multicolumn{1}{c}{\textbf{BLEU-4$\uparrow$}} & \multicolumn{1}{c}{\textbf{WER$\downarrow$}} \\ \midrule
\rowcolor{gray!15}
\multicolumn{9}{l}{\textbf{SALMONN}, \textit{which has \textbf{1600} attention heads in the LLM backbone. }} \\
w/ Instruction & \textbf{2.10} \texttt{|} \textbf{4.95} & \underline{96.79}  & \underline{69.70} & \textbf{93.49} & \underline{20.60} & \underline{40.42} &  \textbf{34.48} & \textbf{23.72}  \\
w/o Instruction & 12.00 \texttt{|} 17.23 &  0.00 & 0.00  & 0.00 & 14.90 & 33.69 & 15.14 & 30.95 \\
w/ random mask & 21.36 \texttt{|} 57.62 & 0.00 & 0.00 & 0.18 & 14.01 & 32.19 & 13.64 & 47.19 \\
w/ \MethodName{} (no instruction) &  \textbf{2.10} \texttt{|} \underline{5.08} &  \textbf{98.05} & \textbf{70.02} & \underline{93.24} & \textbf{24.15} & \textbf{48.71} & \underline{33.90} & \underline{23.89} \\ 
\midrule
\# activated heads in \MethodName{} & 1525 \texttt{|} 1525 & 1259  & 1146 & 1311 & \multicolumn{2}{c}{1426} &  1551 & 1550 \\
\midrule
\midrule
\rowcolor{gray!15}
\multicolumn{9}{l}{\textbf{Qwen2Audio-Instruct}, \textit{which has \textbf{1024} attention heads in the LLM backbone.}}  \\
w/ Instruction & \underline{3.65} \texttt{|} \underline{6.58} &\underline{91.03} & \underline{64.46} & \underline{79.02} & \underline{16.02} & \underline{30.79}& \underline{37.30} & \underline{46.96}\\
w/o Instruction & 217.96 \texttt{|} 234.78  & 0.00 & 0.16 & 0.00 & 12.94 & 26.16 & 0.49 & 85.39 \\
w/ random mask &  164.67 \texttt{|} 200.66 & 0.00 & 0.00 & 0.00 & 13.76 & 28.23& 0.63 & 87.38\\
w/ \MethodName{} (no instruction) & \textbf{2.99} \texttt{|} \textbf{5.65}& \textbf{94.43} & \textbf{70.99} & \textbf{82.34} & \textbf{22.96} & \textbf{46.03} & \textbf{39.02}  & \textbf{46.06}\\
w/ \MethodName{} \scriptsize{of the Base model} &  156.87 \texttt{|} 169.32 &0.00 &  10.23 & 0.02 & 17.53 & 33.80 & 5.89 & 62.28 \\
\midrule
\# activated heads in \MethodName{} & 653 \texttt{|} 649 & 565 & 735 & 527 & \multicolumn{2}{c}{753} &  741 & 721\\
\midrule
\midrule
\rowcolor{gray!15}
\multicolumn{9}{l}{\textbf{Qwen2Audio-Base}, \textit{which has \textbf{1024} attention heads in the LLM backbone.}} \\
w/ Instruction & \textbf{1.77} \texttt{|} \textbf{3.93} & \underline{72.37} & \underline{56.33 }& \underline{49.24} & \underline{12.33} & \underline{27.39}&  \textbf{45.00} & \underline{53.66} \\
w/o Instruction & 99.86 \texttt{|} 99.59 & 0.00 & 0.00 & 0.00 &1.19 & 1.79 & 0.21 & 94.87 \\
w/ random mask & 192.90 \texttt{|} 127.59  & 0.00 & 0.00 & 0.00 &  1.70 & 2.92 &  0.62 & 89.50\\
w/ \MethodName{} (no instruction) & \underline{3.11} \texttt{|} \underline{6.03} & \textbf{89.24} & \textbf{57.05} & \textbf{85.75} & \textbf{21.35} & \textbf{44.89} & \underline{39.27} & \textbf{38.37}\\
w/ \MethodName{} \scriptsize{of the Instruct model} & 252.96 \texttt{|} 350.77 & 0.00 & 0.00 & 0.00 & 2.99 &5.41 & 12.73 & 64.55 \\
\midrule
\# activated heads in \MethodName{} & 726 \texttt{|} 724 & 633 & 679 & 797 & \multicolumn{2}{c}{729} & 769 & 774 \\
\bottomrule
\end{tabular}
}
\vspace{-0.03in}
\caption{Single task performance of SALMONN, Qwen2Audio-Instruct, and Qwen2Audio-Base. WERs in the ASR task are formatted as LibriSpeech test-clean\texttt{|}test-other. When applying \MethodName{}, no instruction is given. When applying a random mask, we ensure the number of activated heads is the same with the corresponding \MethodName{} in that task. Note that the number of trainable parameters in \MethodName{} is the same as number of attention heads in the LLM backbone. 
}
\vspace{-0.1in}
\label{tab:single-salmonn}
\end{table*}

\begin{table*}[t]
\centering
\small{
\begin{tabular}{@{}lccccccccc@{}}
\toprule
\textbf{Task} & \multicolumn{3}{c}{\textbf{GR+``\texttt{|}''+ASR}} & \multicolumn{3}{c}{\textbf{ASR+``\texttt{|}''+GR}} & \multicolumn{3}{c}{\textbf{Json style \{``ASR'': , ``GR'': \}}} \\
\cmidrule(lr){2-4}
\cmidrule(lr){5-7}
\cmidrule(lr){8-10}
\textbf{Metric} & \textbf{IFR}$\uparrow$ & \textbf{ACC}$\uparrow$ & \textbf{WER}$\downarrow$ & \textbf{IFR}$\uparrow$ & \textbf{WER}$\downarrow$ & \textbf{ACC}$\uparrow$ & \textbf{IFR}$\uparrow$ & \textbf{WER}$\downarrow$ & \textbf{ACC}$\uparrow$ \\
\midrule
\rowcolor{gray!15}
\multicolumn{10}{l}{\textbf{SALMONN}, \textit{which has \textbf{1600} attention heads in the LLM backbone.}} \\
w/ Instruction & 98.59 & 68.02 & 3.52 & 16.03 & 29.36 & 45.95 & 69.16 & 6.17 & 51.05 \\
w/ \MethodName{} (no instruction) & \textbf{99.12} & \textbf{97.77} & \textbf{2.21} &  \textbf{97.63} & \textbf{2.29} &  \textbf{97.81}& \textbf{98.89} & \textbf{2.40} & \textbf{97.30}  \\
\midrule
\# activated heads in \MethodName{} & \multicolumn{3}{c}{1315}  & \multicolumn{3}{c}{1252}  &  \multicolumn{3}{c}{1124}   \\
\midrule
\midrule
\rowcolor{gray!15}
\multicolumn{10}{l}{\textbf{Qwen2Audio-Instruct}, \textit{which has \textbf{1024} attention heads in the LLM backbone.}} \\
w/ Instruction & 47.44 & 68.79 & 7.03& 79.16 & 50.34& 53.09 & 0.00 & - & - \\
w/ \MethodName{} (no instruction) & \textbf{94.62} & \textbf{91.09}& \textbf{3.08}&\textbf{89.92} &\textbf{3.22} & \textbf{89.56} & \textbf{58.45} & \textbf{3.87} & \textbf{89.16}\\
\midrule
\# activated heads in \MethodName{} & \multicolumn{3}{c}{555} & \multicolumn{3}{c}{558} & \multicolumn{3}{c}{522} \\ 
\midrule\midrule
\rowcolor{gray!15}
\multicolumn{10}{l}{\textbf{Qwen2Audio-Base}, \textit{which has \textbf{1024} attention heads in the LLM backbone.}} \\
w/ Instruction & 13.74 & 1.11 & 58.87 & 10.92 & 68.75 & 4.20 & 0.00 & - & - \\
w/ \MethodName{} (no instruction) & \textbf{82.40} & \textbf{64.94}&  \textbf{3.25}& \textbf{33.93} &\textbf{6.51} &\textbf{58.83} & \textbf{71.19} & \textbf{2.79} & \textbf{87.46} \\
\midrule
\# activated heads in \MethodName{} & \multicolumn{3}{c}{641} & \multicolumn{3}{c}{599} & \multicolumn{3}{c}{616} \\
\bottomrule
\end{tabular}
}
\vspace{-0.05in}
\caption{Composite task performance of SALMONN, Qwen2Audio-Instruct, and Qwen2Audio-Base. IFR denotes instruction following rate (\%). Task-specific metrics are calculated among instruction-following samples. 
Note that the trainable parameter count in \MethodName{} is the same as number of attention heads in the LLM backbone.
}
\label{tab:composite-salmonn}
\vspace{-0.1in}
\end{table*}

\subsection{Prompt Sensitivity in LALMs}

\citet{peng2024survey} point out that speech LLMs exhibit non-negligible sensitivity to linguistic variations in the prompts.
Following \citet{peng2024survey} and \citet{sclar2024quantifying}, we present more evidence to showcase this issue for general LALMs.
For each task, we first write a \textbf{general} instruction, and then apply one of the following variations:
\begin{itemize}
    \item \textbf{Alter symbol} and \textbf{Alter case}, where we only change punctuation marks or letter casings.
    \item \textbf{Semantically equal (simple, neutral, complex)}, where the general prompt is rewritten with the same semantic meaning but in short, medium, or long sentences. This is done by providing Gemini 2.5 Flash with manual rephrasing examples and asking it to generate more.
    \item When the task is classification (GR, SER, ASV), we also apply \textbf{Add constraint} where the set of possible classes are provided in the prompt. For example, ``Happy, Sad, Angry, or Neutral?'' for SER.
\end{itemize}
For \textbf{Alter symbol}, \textbf{Semantically equal} and \textbf{Add constraint} variations, we construct 5 different prompts in each type. 
We take SALMONN as a typical LALM here.
Notably, all 7 individual tasks here are comprehensively covered within the training data of SALMONN.
Our preliminary experiments indicate that SALMONN demonstrates greater robustness to variations in instructions compared to Qwen2Audio.
For decoding, we use deterministic beam search with beam size 4, so that the difference in model performance is only caused by different instructions.

The performance with instruction variations are visualized in Fig. \ref{fig:sensitivity}. 
These results clearly indicate that SALMONN still has high sensitivity to instructions with the same intention.
For example, in ASR task, changing the prompt to all capital letters will cause more hallucinations (repeating or generating upper case phonemes instead of words), thus raising the WER to 12\%.
In GR task, adding the constraint ``Male or Female'' makes the model almost always output ``Male''.
Similar phenomenon can also be observed with SER.
For SALMONN, ASV and S2TT tasks exhibit better robustness to instruction variations.
In almost all cases, using longer and more complex instructions is likely to cause a degradation in performance, sometimes even drastically as in ASR and SER.
These results exemplify that prompt sensitivity remains a significant challenge in LALMs.

\begin{figure*}[t]
    \centering
    \includegraphics[width=0.3\linewidth]{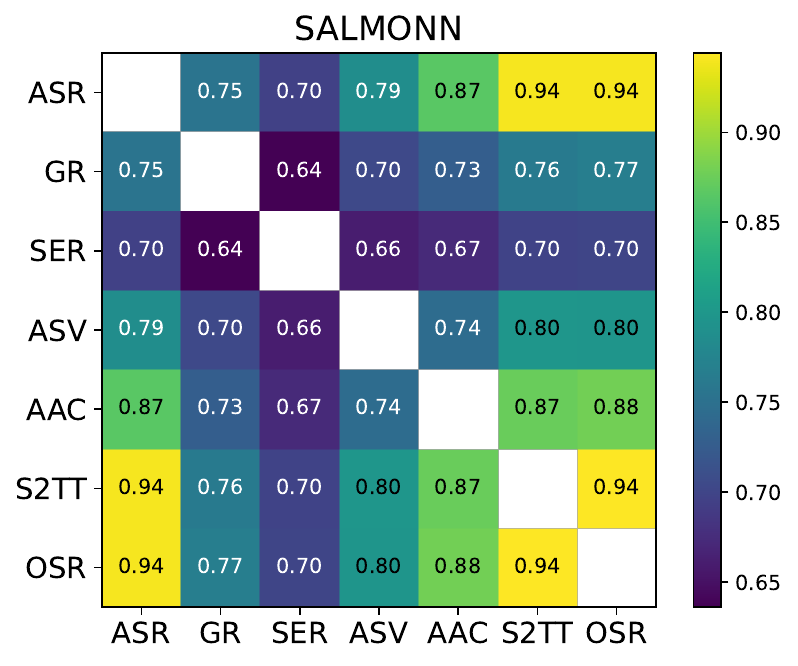}
    \includegraphics[width=0.3\linewidth]{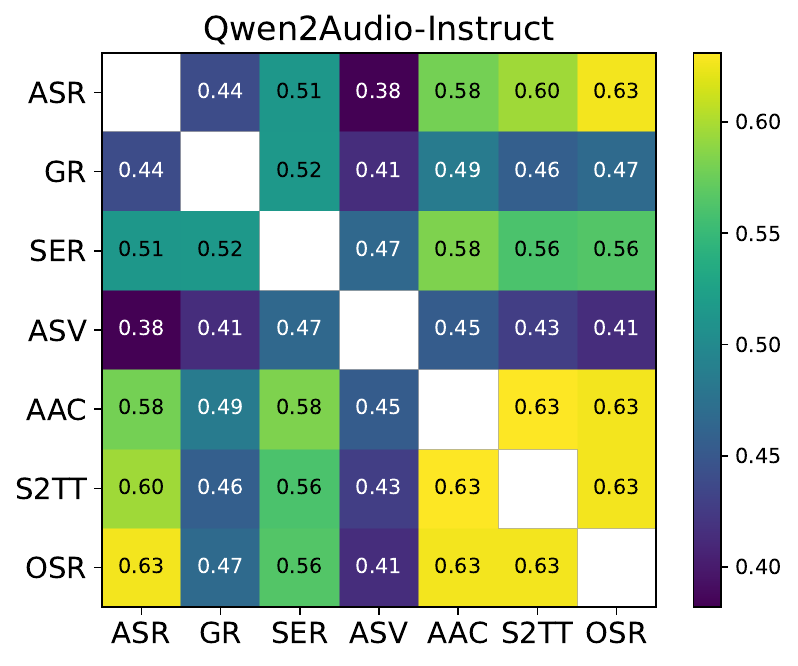}
    \includegraphics[width=0.3\linewidth]{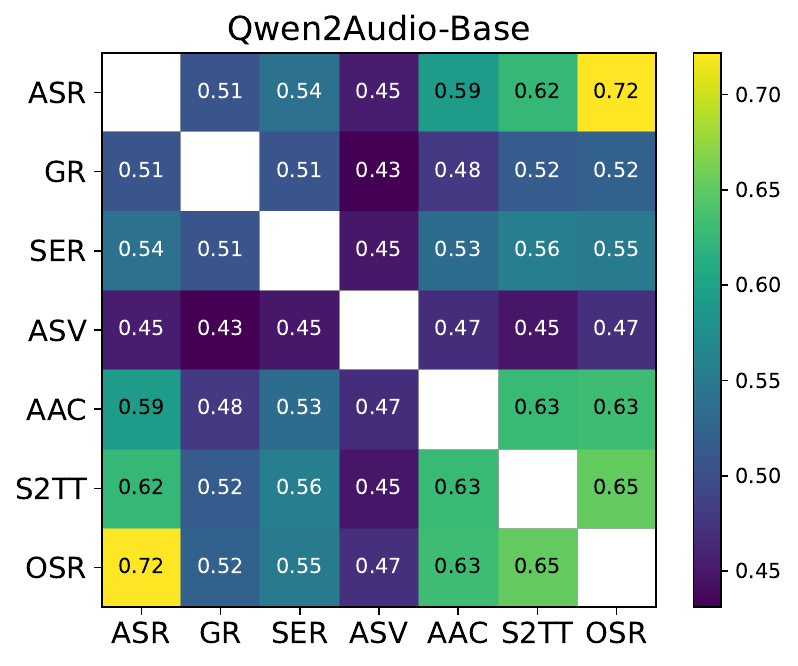}
    \vspace{-0.1in}
    \caption{Jaccard similarities of \MethodName{} between different tasks in each LALM.}
    \label{fig:jaccard-sim}
    \vspace{-0.12in}
\end{figure*}

\subsection{\MethodName{} for Single Tasks}
We apply \MethodName{} on SALMONN and Qwen2Audio variants, on all the 7 single tasks.
We compare LALM performance with our method (``w/ \MethodName{}'') against three other methods: \textcircled{1} with clear instructions (w/ Instruction) from the ``general'' type in the previous section, \textcircled{2} without instructions (w/o Instruction), and \textcircled{3} using a random mask with the same number of activated heads as \MethodName{} for each task.
For Qwen2Audio, since the two variants are identical in structure, we also evaluate \textcircled{4} swapped \MethodName{} across variants.

The results in Table \ref{tab:single-salmonn} indicate several insights:
\begin{itemize}
    \item LALMs with only certain attention heads activated can achieve comparable or even better performance than using natural language instructions on all of the 7 single audio understanding tasks. 
    Even though LALMs with natural language instructions sometimes still outperform those with only \MethodName{}, the difference in performance is very small (e.g. on ASV, S2TT and OSR tasks with SALMONN).
    Note that AHAMask is only supposed to reliably \textbf{manifest} specific functionalities instead of \textbf{improving} them, so these are still expected and supporting evidence.
    Only on ASR and S2TT with Qwen2Audio-Base, which are the pretraining tasks of this model, using instructions achieves noticeably better performance compared to \MethodName{}.
    \item Even on tasks where the original LALM struggles to perform well with explicit instructions, applying \MethodName{} can trigger its functionality to perform much better.
    This is especially evident for Qwen2Audio-Base, which cannot perform well on tasks that are covered by little or no data in the pretraining phase.
    Applying \MethodName{} on this base model significantly improves its performance on GR, ASV, ACC and OSR to match or even surpass the instructed model Qwen2Audio-Instruct.
    \item An \MethodName{} is only valid at specific head positions; randomly shuffling it will fail, as indicated by the ``w/ random mask'' results.
    \item An \MethodName{} is only valid for the model it is trained on; swapping from different variants under the same model family will not work, as for Qwen2Audio. 
    \item It does not require dropping many attention heads to achieve reliable task specification. For example, only masking 75 and 49 out of 1600 heads is enough for SALMONN to perform ASR and S2TT without instructions, respectively.
    \item The number of activated heads in \MethodName{} seems to correlate with task complexity. For simpler classification tasks like SER, GR and ASV, it usually requires fewer heads; for other sequence generation tasks, more attention heads are activated. However, the base model Qwen2Audio-Base exhibits different characteristics compared to Qwen2Audio-Instruct on this property.
\end{itemize}

\subsection{\MethodName{} for Composite Tasks}
From the previous section, it is clear that LALMs can perform single acoustic tasks without instructions once equipped with task-specific \MethodName{}s.
In this section, we evaluate the effect of \MethodName{} on more complex tasks.
We design composite multi-hop tasks that require LALMs to generate answers in a specific format, hence also bring the metric of instruction following rate (IFR).
Here, we consider asking the model to perform two tasks in order with a separator ``\texttt{|}'', or in json format with keys ``ASR'' and ``GR''.
IFR is computed as the percentage where the generated answers can be divided into two parts by ``\texttt{|}'', or parsed by json.
We combine the ASR and GR tasks as they are largely independent of each other, thus require LALMs to attend to different acoustic aspects.
We use LibriSpeech for this experiment.

We present the results in Table \ref{tab:composite-salmonn}.
In all the metrics, applying \MethodName{} achieves better performance than using instructions, especially for Qwen2Audio-Base where it is hard to instruct the model well.
Also, LALMs can be sensitive to the ordering of sub-tasks in such composite tasks, e.g. the original SALMONN model can generate answers in the format of ``GR\texttt{|}ASR'' well, but fails in the opposite ordering.
Even if the specific format is satisfied, the performance in each sub-task is not as good as when only doing single tasks.
For SALMONN and Qwen2Audio-Instruct, the gap between different task orders is largely reduced using \MethodName{}, and the performance of each sub-task is closer to the single task scenarios.
These results further indicate that \MethodName{} can control  more complex behavior of LALMs in processing multiple aspects of acoustic information.


\subsection{Analysis of Acoustic Attention Head Masks}

\begin{figure}
    \centering
    \includegraphics[width=0.99\linewidth]{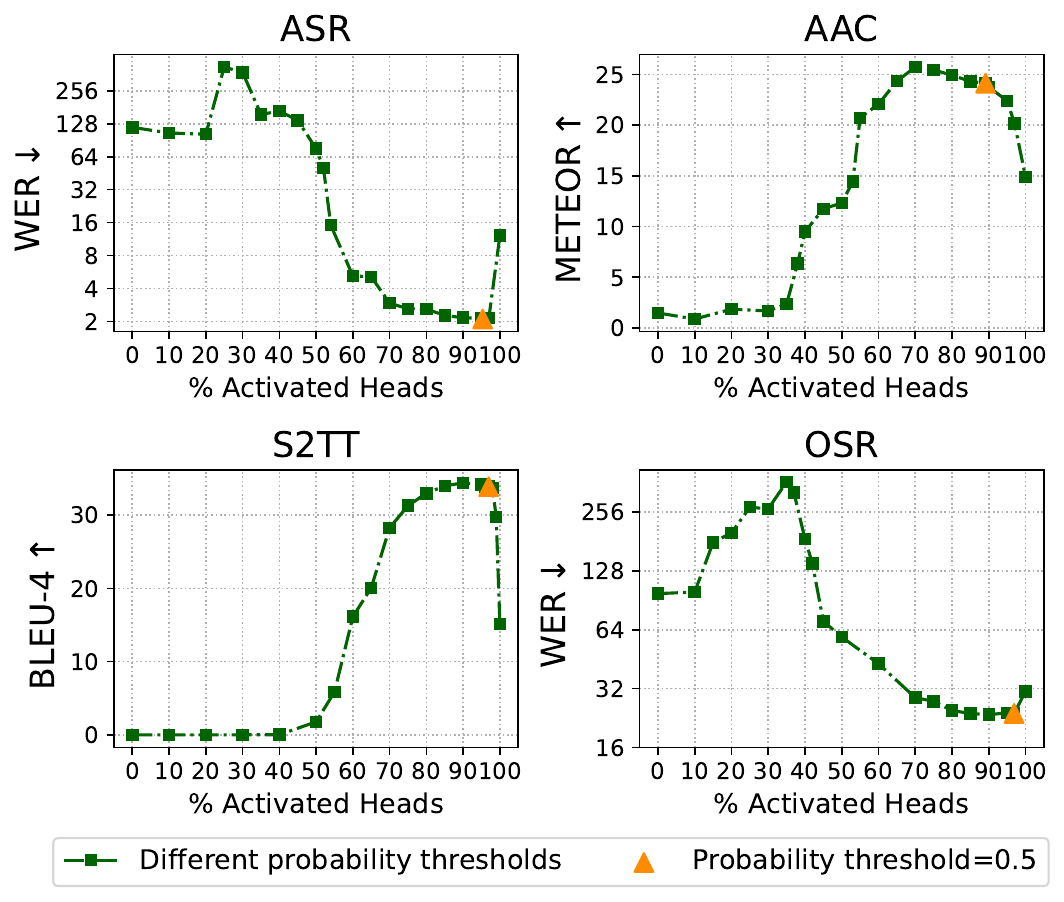}
    \caption{Performance of SALMONN on 4 non-classification tasks with \MethodName{} in different percentage of activated attention heads. 
    The orange triangle marker denotes the metric in Table \ref{tab:single-salmonn}
    }
    \label{fig:percentage-heads}
    \vspace{-0.1in}
\end{figure}
\subsubsection{The Similarity of Masks for Different Tasks}
Intuitively, the relation between the sets of activated attention heads should be correlated to the corresponding tasks.
More similar tasks should come with more common attention heads.
From previous results, we can already observe that the number of attention heads has a correlation with task complexity.
In this section, we visualize the similarity of \MethodName{} between different tasks in all the three LALMs.
We consider Jaccard similarity to measure the similarity of two boolean sets: $J(\mathcal M_1,\mathcal M_2)=|\mathcal M_1\cap\mathcal M_2|/|\mathcal M_1\cup \mathcal M_2|$.

Fig.\ref{fig:jaccard-sim} visualizes the Jaccard similarity matrices. 
As expected, tasks that are intuitively similar share more attention heads.
For example, \MethodName{}s for OSR and ASR tasks are the most similar in all three LALMs, while paralinguistic and classification tasks like SER and ASV usually have lower similarity with others.
Interestingly, the AAC task also has good similarity in \MethodName{} with ASR, S2TT and OSR tasks.
The similarity matrices for different LALMs also exhibit different patterns.
These mask similarities very likely stem from LALMs' internal mechanisms, which may inspire future work on interpreting their behaviors.

\subsubsection{Attention Heads Gradually Form Acoustic Functionalities}

Since we essentially train a logit matrix $\mathbf M$, each attention head is associated with a continuous mask probability in $\sigma(\mathbf M)$.
We can then apply different thresholds to these probabilities to obtain masks with different percentage of activated attention heads. 
This will indicate if there are gradual or abrupt changes in the performance when we activate the attention heads in order.
In other words, here we explore how the acoustic functional pathways of LALMs are formed by the attention heads.

To achieve a desired activation percentage $q$ of heads, we mask the attention heads whose corresponding logits in $\mathbf M$ fall within the lowest $1-q$ quantile.
We do so for different percentage levels on all the single tasks for SALMONN.
However, as the accuracy metrics in classification tasks (GR, SER, ASV) treat all wrong classifications equally, they are not sufficient to tell the evolution of functionalities.
Hence, we only show the results on non-classification tasks in Fig. \ref{fig:percentage-heads}, where the metrics are more fine-grained.
It is thus clear that many attention heads have an incremental improvement on the task functionality, and the optimal metrics are obtained based on collective efforts.

For classification tasks, we also show in Table \ref{tab:critical-heads-example} an example of SALMONN producing different types of outputs with different number of activated heads in the order of head weights.
Before the 848-th head, the model only generates either random or empty strings.
After the 848-th head, it is able to generate emotion descriptors, but not correct ones, until the 873-rd head.
After the 1429-th head, it loses classification ability, and the model starts to drift away to emotional cue (``crying''), and finally plain transcriptions.
These results indicate that LALM's acoustic functional pathway is gradually instead of suddenly formed based on the importance of heads, which provides additional insights to studying the explainability of such models.


\begin{table}[]
\centering
\small{
\begin{tabular}{@{}lll@{}}
\toprule
\multicolumn{1}{c}{\textbf{\# Activated Heads}} & \multicolumn{1}{c}{\textbf{Output}} & \multicolumn{1}{c}{\textbf{Output Type}} \\ \midrule
320 (20\%) & at 0) - 0 0) 0) 0) 0 \&  & \multirow{2}{*}{\begin{tabular}[c]{@{}l@{}}Meaningless\\Strings\end{tabular}} \\
847 (52.9375\%) & \textit{(empty string)}\\
\midrule
848 (53\%) & Happy & \multirow{4}{*}{\begin{tabular}[c]{@{}l@{}}Incorrect\\Emotions\end{tabular}}\\
862 (53.875\%) & Happy \\
863 (53.9375\%) & Neutral \\ 
872 (54.5\%) & Neutral \\ 
\midrule
873 (54.5625\%) & Sad & \multirow{2}{*}{\begin{tabular}[c]{@{}l@{}}Correct\\Emotions\end{tabular}}\\ 
1429 (89.3125\%) & Sad \\ 
\midrule
1430 (89.375\%) & Someone is crying. & \multirow{4}{*}{\begin{tabular}[c]{@{}l@{}}Captions or\\Transcriptions\end{tabular}}\\ 
1579 (98.6875\%) & Someone is crying. \\ 
1580 (98.75\%) & I know. \\
1593 (99.5625\%) &I  know, I know.\\
\bottomrule
\end{tabular}
}
\vspace{-0.05in}
\caption{The output of SALMONN on one example utterance in the SER task, with different number of activated heads. 
The heads are activated in the order of their weight in $\mathbf M$.
The original utterance contains a female crying ``I know, I know.'' with ground truth emotion ``Sad''.
}
\label{tab:critical-heads-example}
\vspace{-0.1in}
\end{table}

\section{Conclusion}
In this work, we propose \MethodName{} that simply masks some attention heads in an LALM to address the prompt sensitivity problem and achieve reliable task specification without instructions.
These masks are efficiently obtained through a lightweight training process.
We verify by experiments that applying \MethodName{} can achieve comparable or even better performance than using general instructions, in diverse scenarios. 
Our findings also suggest that acoustic ``functional pathways'' do exist in the attention heads of LALMs, which can be an important inspiration for the interpretability of such models.
Future research includes analyzing the partitioning and composability of \MethodName{} and constructing a general text-to-mask converter for more use cases.


\section*{Acknowledgements}
This work is supported by Open Project of the Key Laboratory of Artificial Intelligence, Ministry of Education (No. AI202405), and the China NSFC Project (No. 92370206).

\bibliography{aaai2026}
\appendix

\section{Appendix A: Data and Model Details}

The training and test data sizes and example outputs for each task can be seen in Table \ref{tab:dataset-more}, together with detailed information about LALMs.
We use official checkpoints for all LALMs.
All waveforms are downsampled to 16kHz before feeding the LALMs.
For SALMONN, the input waveforms are padded or truncated to 30s, and both its Whisper~\cite{whisper} and BEATs~\cite{beats} encoders output high dimensional embedding sequences at 50Hz frame rate.
Then, a Q-former~\cite{li2023blip} is applied to summarize the acoustic sequences into 88 embeddings for the 30s audio segment before feeding the LLM backbone.
For Qwen2Audio models, its encoder compresses 100Hz spectrogram inputs into  25Hz embedding sequences before the LLM backbone.
For Qwen2Audio-Instruct, the official chat template is also applied on the audio and text data to enable user-assistant dialogue turns.
For SALMONN and Qwen2Audio-Base, the instructions tokens are simply appended after the audio features, such as ``\texttt{<|audio\_bos|><|audio|><|audio\_eos|> <|instruction|> <|output|>}''

For both SALMONN and Qwen2Audio, we use float16 dtype in training.
For SALMONN, we set batch size to 8 for all tasks, except 6 for OSR and 4 for S2TT because of the longer sequence lengths, and train for at most 90k steps.
For Qwen2Audio, we set batch size to 4 for all tasks, and train for at most 300k steps.
We observe that training AHAMask on SALMONN converges faster and exhibits less oscillation than Qwen2Audio, probably because of the larger model size (13B compared to 7B in the LLM backbone).

In decoding, for SALMONN, we use deterministic beam search with beam size 4 to follow its official configuration; for Qwen2Audio variants, we use greedy search (beam size 1). 
This ensures no randomness in decoding.

\begin{figure*}
    \centering
    \includegraphics[width=0.99\linewidth]{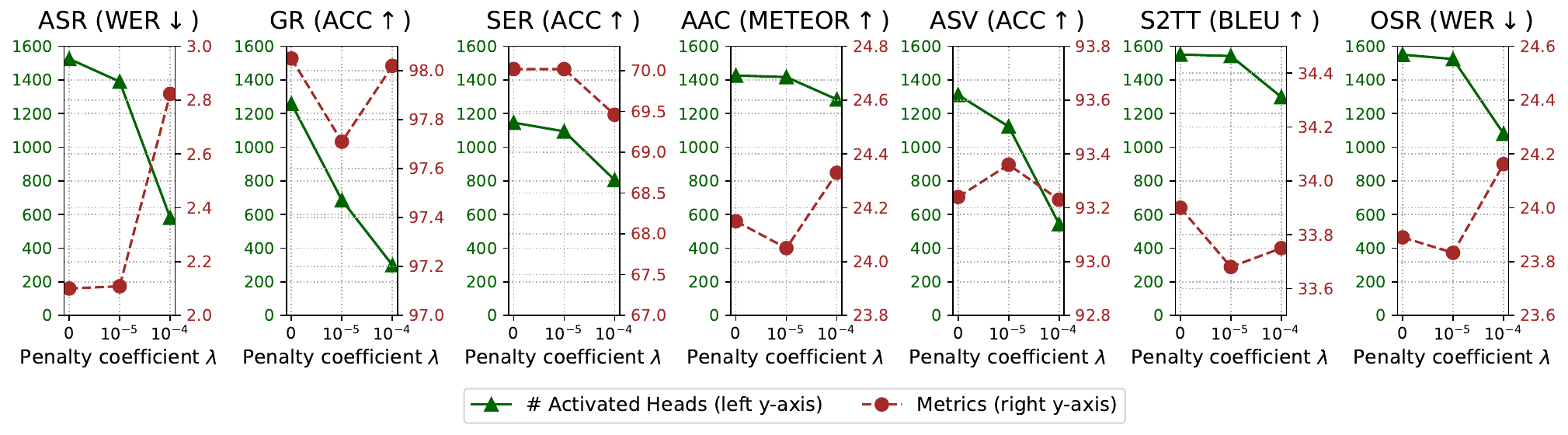}
    \caption{The effect of the penalty coefficient $\lambda$ on the number of activated attention heads on SALMONN. Larger $\lambda$ leads to fewer activated heads (green solid lines), but not necessarily worse performance (red dashed lines). Note that the difference in performance caused by different $\lambda$ is usually small, i.e. the right y-axis usually has a small range.}
    \label{fig:lambda}
\end{figure*}

\section{Appendix B: More Explorations of AHAMask}

\begin{table}[]
\centering
\resizebox{\linewidth}{!}{
\begin{tabular}{@{}cccc@{}}
\toprule
\textbf{Task} & \textbf{Metric} & \textbf{w/ Instr.} & \textbf{w/ AHAMask} \\ \midrule
\multirow{3}{*}{SER+``\texttt{|}''+GR} & IFR$\uparrow$ & 99.84 & 100.00 \\
 & SER ACC$\uparrow$ & 45.28 & 64.46 \\
 & GR ACC$\uparrow$ & 48.10 & 97.18 \\\midrule
\multirow{3}{*}{GR+``\texttt{|}''+SER} & IFR$\uparrow$ & 100.00 & 100.00 \\
 & GR ACC$\uparrow$ & 80.98 & 97.50 \\
 & SER ACC$\uparrow$ &   40.45  & 64.71\\\midrule
\multirow{3}{*}{\{``SER'':~,``GR'':~\}} & IFR$\uparrow$ & 87.99 & 95.89\\
 & SER ACC$\uparrow$ &  30.86 & 67.23 \\
& GR ACC$\uparrow$ &  47.44 & 95.71 \\\bottomrule
\end{tabular}
}
\caption{Performance of AHAMask on composite emotion and gender recognition tasks.}
\label{tab:composite-ser-gr}
\end{table}

\begin{table}[]
\centering
\begin{tabular}{@{}cccc@{}}
\toprule
\textbf{Task} & \textbf{Metric} & \textbf{w/ Instr.} & \textbf{w/ AHAMask} \\ \midrule
\multirow{2}{*}{GR+``\texttt{|}''+ASR} & ACC$\uparrow$ & 67.33 & 97.25 \\
 & WER$\downarrow$ & 3.56 &  3.63\\ \midrule
\multirow{2}{*}{ASR+``\texttt{|}''+GR} & WER$\downarrow$ & 6.10 & 3.31 \\
 & ACC$\uparrow$ &  12.25 &  95.95 \\ \midrule
\multirow{2}{*}{\{``ASR'':~,``GR'':~\}} & WER$\downarrow$ & 7.69 & 3.71 \\
 & ACC$\uparrow$ & 36.83  & 96.41 \\ \bottomrule
\end{tabular}
\caption{Evaluation of single task performance on SALMONN under composite scenarios and LLM-corrected formats. The setup is similar to Table \ref{tab:composite-salmonn}, which does not employ an LLM to repair the LALM outputs. We omit the IFR metric here because it is 100\% after LLM format correction. WER and ACC are for ASR and GR tasks, respectively. ``Instr.'' means ``instructions''.}
\label{tab:LLM-corrected}
\end{table}

In this section, we show more experimental results of AHAMask on SALMONN.

\subsection{Additional Composite Tasks}
We also evaluate AHAMask on other composite tasks besides ASR and GR. 
Table \ref{tab:composite-ser-gr} shows the results on SER and GR tasks with IEMOCAP training and test sets.
These results are similar to Table \ref{tab:composite-salmonn}, which provide stronger evidence on the validity of AHAMask.

\subsection{Performance of Single Tasks under Composite Scenarios}
In Table \ref{tab:composite-salmonn}, we evaluated LALMs on composite tasks, either with instructions or AHAMask. In that comparison, we calculated instruction following rates, and the metrics on single tasks only within instruction-following samples.
Here, we consider using an LLM to reconstruct the required format for those instruction non-following samples, and measure the performance of single tasks on all the samples.
This LLM is instructed to read the LALM outputs that do not follow the required format, infer the outputs for each single task, and assemble them into the required format.
When a task is missing in the LALM ouput, the LLM is asked to use an empty string for that task.
We use OpenAI \texttt{gpt-4o-mini} API which exhibits 100\% instruction following rate on the tested samples.
Then, we obtain the task-specific metrics in Table \ref{tab:LLM-corrected}.

These results show that the advantage of AHAMask still remains with LLM format corrections.
Compared to Table \ref{tab:composite-salmonn}, all metrics become worse, because certain tasks are usually missing when the format is not satisfied.
For the GR+``\texttt{|}''+ASR composite task, the ASR WER of AHAMask is slightly higher than that of official SALMONN with instructions.
We identify that it is because of more deletion errors, i.e. the ASR task is usually missing among the 0.88\% instruction non-following samples from AHAMask.

\subsection{On the Generalizability of AHAMask}
\begin{table}[]
\centering
\begin{tabular}{@{}lcc@{}}
\toprule
\multicolumn{1}{c}{\textbf{Test Set}} & \multicolumn{1}{c}{\textbf{w/ Instruction}} & \multicolumn{1}{c}{\textbf{w/ AHAMask}} \\ \midrule
\begin{tabular}[c]{@{}l@{}}LibriSpeech \small{(test-clean)}\end{tabular} &  96.79 & 98.05 \\ \midrule
\begin{tabular}[c]{@{}l@{}}IEMOCAP \small{(Session 5)}\end{tabular} & 88.64 & 91.62 \\ 
\begin{tabular}[c]{@{}l@{}}TEDLIUM \small{(test)}\end{tabular} & 93.83 & 95.65 \\ 
\begin{tabular}[c]{@{}l@{}}CommonVoice11 \small{(test)}\end{tabular} &  50.73 & 89.96 \\ 
\begin{tabular}[c]{@{}l@{}}VoxCeleb1 \small{(test)}\end{tabular} & 95.90 & 98.75 \\ \bottomrule
\end{tabular}
\caption{Out-of-domain evaluation on GR task on SALMONN. Metric is ACC (\%). Parentheses denote dataset partition. Note that we use the same AHAMask trained on LibriSpeech train-clean-100 in the last column.}
\label{tab:ood-gr}
\end{table}

In almost all experiments in this paper, each AHAMask is obtained by training only on one dataset.
Ideally, since each AHAMask is supposed to activate a function of the LALM, it should generalize well across different domains.
However, as it is acquired by data-driven neural network training in practice, its generalizability also needs verification.

We first observe that for SALMONN, on the GR task, AHAMask exhibits good generalizability on out-of-domain data, despite only being trained on LibriSpeech.
Results in Table \ref{tab:ood-gr} demonstrates that the AHAMask for GR task on SALMONN performs well on IEMOCAP, TEDLIUM~\cite{rousseau2012tedlium}, CommonVoice11, and VoxCeleb1 test sets.

Then, on ASR task, we find the generalizability of AHAMask still has room of improvement.
When evaluating on the out-of-domain TEDLIUM test set, we observe noticeable discrepancy between AHAMask (trained on LibriSpeech only) and the official instructed model.
When we use mixed training set for AHAMask, such as a balanced mix from LibriSpeech and CommonVoice11, we obtain WER scores that match well with the instructed model.
Those results are shown in Table \ref{tab:ood-asr}.
This may indicate that AHAMask can still be affected by the training data domains.
When trained on a single domain, it may capture a more fine-grained functionality (such as \textit{transcribing clean audiobook recording}).
Therefore, using more diverse data domains, like the way original LALMs are trained, is always beneficial.
It remains a future work on evaluating and improving the generalizability of AHAMask, so that it can manifest the general functionality of LALMs more steadily.

\begin{table}[]
\centering
\begin{tabular}{@{}lcc@{}}
\toprule
\multicolumn{1}{c}{\textbf{AHAMask Training Data}} & \multicolumn{1}{c}{\textbf{w/ AHAMask}}  & \multicolumn{1}{c}{\textbf{w/ Instr.}} \\ \midrule
LibriSpeech (all, 280k) & 4.01  & \multirow{3}{*}{3.65} \\
 \cmidrule(r){1-2}
\begin{tabular}[c]{@{}l@{}}LibriSpeech (50k)\\ ~~+CommonVoice11 (50k)\end{tabular} & 3.67  & \\ \bottomrule
\end{tabular}
\caption{Out-of-domain evaluation on ASR task on SALMONN. Metric is WER (\%). Parentheses denote the number of training samples. The test set here is TEDLIUM (test), which is unseen for both the original SALMONN model and AHAMask. ``Instr.'' means ``instructions''.}
\label{tab:ood-asr}
\end{table}

\subsection{Reducing the Number of Activated Heads}

Previous results in the main text have shown that, in many cases, only masking tens of attention heads can be enough to achieve comparable performance against using natural language instructions.
For example, only masking 50 out of 1600 heads can trigger OSR functionality.
In this section, we argue that there are redundant attention heads, and we can further reduce the number of necessary attention heads for a specific task without compromising performance.

To explicitly put a constraint on the number of activated attention heads during training, we can apply a penalty loss term.
Suppose $\lambda$ is the loss weight for this penalty loss.
The overall loss for training $\mathcal M$ is then 
\begin{equation}
    \mathcal L=\mathcal L_{\text{CE}}+\lambda \sum_{i,j} m_{i,j}
\end{equation}
where $\mathcal L_{\text{CE}}$ is the original next-token prediction loss, $ m_{i,j}\in \mathcal M$ is the binary indicator for each head after Gumbel-Sigmoid. 
The value of $\lambda$ is 0 in previous experiments.
In this experiment, we change $\lambda$ to $10^{-5}$ and $10^{-4}$ for SALMONN.
Considering SALMONN has 1600 heads in its LLM backbone, setting $\lambda=10^{-4}$ will introduce a penalty loss around $0.1$, which is already a strong supervision signal.
The resulting number of activated attention heads and corresponding performance on all the 7 single tasks can be seen in Fig. \ref{fig:lambda}.
This suggests that, in many tasks, reducing the number of activated heads does not lead to much performance degradation.
For example, SALMONN can achieve 98.02\% accuracy on GR using only 299 out of 1600 heads, while it reaches only 98.05\% by 1259 heads even without any constraints.
It is only on ASR task that reducing the number of heads to 581 results in significantly higher WER.
Moreover, the change of model performance along with the increase of $\lambda$ is not monotonic.
Sometimes masking more attention heads is beneficial.
This indicates it is possible that some attention heads have negative effects at a specific task.

\subsection{The \textit{All Roads Lead to Rome} Effect}

It has been shown that certain attention heads in the LLM backbone of an LALM can form acoustic pathways. 
We also find that these pathways are not unique; 
in fact, there exists an \textit{all roads lead to Rome} effect.
There are different multiple sets of attention heads that can reach the same acoustic functionality equally well.
To see this, we train AHAMask on SALMONN on the GR and SER task using three random seeds.
Then we run inference on the corresponding test data using these masks under the same criterion.
We do the trainings with $\lambda=10^{-4}$ so the number of activated heads is already restricted.

The results are presented in Table \ref{tab:rome-effect}.
It is then clear that two AHAMasks can achieve similar GR or SER performance even if they are very different. 
Note that although different random seeds may cause a fluctuation in performance, this variation is magnitudes smaller than the sensitivity when using instructions (i.e., compared to  Fig. \ref{fig:sensitivity}).
This means the acoustic functional pathway for a specific task can be constructed in multiple ways that are equally well.

An interesting application of this effect is to perform boolean operation on these masks, to further reduce the number of activated heads.
For example, we can perform AND operation on all the three different AHAMasks for each task, so as a head is only activated when it is activated in all masks.
We show this result in Table \ref{tab:rome-effect} as well.
Surprisingly, even if the number of activated heads is further reduced, the performance of such 3-way-AND AHAMask does not degrade.
Instead, the performance of the intersected mask lies in the middle of those of all the three masks.
Now for GR, we can use only 230 heads to achieve 98.05\% accuracy.
This technique of intersecting masks may help us find the smallest set of necessary heads for an acoustic task, which may be used for more interesting explorations in future work.

\begin{table}[]
\centering
\small{
\begin{tabular}{@{}llccc@{}}
\toprule
\multicolumn{1}{c}{\textbf{Task}} & \multicolumn{1}{c}{\textbf{AHAMask}} & \multicolumn{1}{c}{\textbf{ACC$\uparrow$}} & \multicolumn{1}{c}{\textbf{\begin{tabular}[c]{@{}c@{}}Activated\\ Heads\end{tabular}}} & \multicolumn{1}{c}{\textbf{\begin{tabular}[c]{@{}c@{}}Diff. Ratio\\ to $\mathcal M_1$\end{tabular}}} \\ \midrule
\multirow{4}{*}{GR} & $\mathcal M_1$ (seed 42) & 98.02 & 299 & - \\
 & $\mathcal M_2$ (seed 43) & 98.28 & 319 & 32.1\% \\
 & $\mathcal M_3$ (seed 44) & 97.90 & 300 & 33.1\% \\ \cmidrule(l){2-5} 
 & $\bigcap_{i=1}^3 \mathcal M_i$ & 98.05 & 230 & 23.1\% \\ \midrule
\multirow{4}{*}{SER} & $\mathcal M_1$ (seed 42) & 69.46 & 805 & - \\
 & $\mathcal M_2$ (seed 43) & 70.59 & 790 & 10.8\% \\
 & $\mathcal M_3$ (seed 44) & 68.41 & 787  & 12.7\% \\ \cmidrule(l){2-5} 
 & $\bigcap_{i=1}^3 \mathcal M_i$ & 69.64 & 723 & 10.2\% \\
 \bottomrule
\end{tabular}
}
\caption{An example of the \textit{all roads lead to Rome} effect on SALMONN on GR and SER tasks.
We obtain $\mathcal M_1,\mathcal M_2,\mathcal M_3$ using three different random seeds in training.
``Diff. Ratio to $\mathcal M_1$'' is calculated as $|\mathcal M_i\ne\mathcal M_1|/|\mathcal M_1|$ for $i=2,3$.
The row $\bigcap _{i=1}^3 \mathcal M_i$ means we only activate heads that are activated in all $\mathcal M_1,\mathcal M_2,\mathcal M_3$.
}
\label{tab:rome-effect}
\end{table}


\section{Appendix C: Used Instructions}
For reproducibility, please refer to Table \ref{tab:instructions} for detailed instructions used in the main text.
Please also refer to the supplementary files for detailed instructions we use for SALMONN in the instruction sensitivity experiment.


\section{Appendix D: Visualization of AHAMasks}
\begin{figure*}
    \centering
    \includegraphics[width=0.99\linewidth]{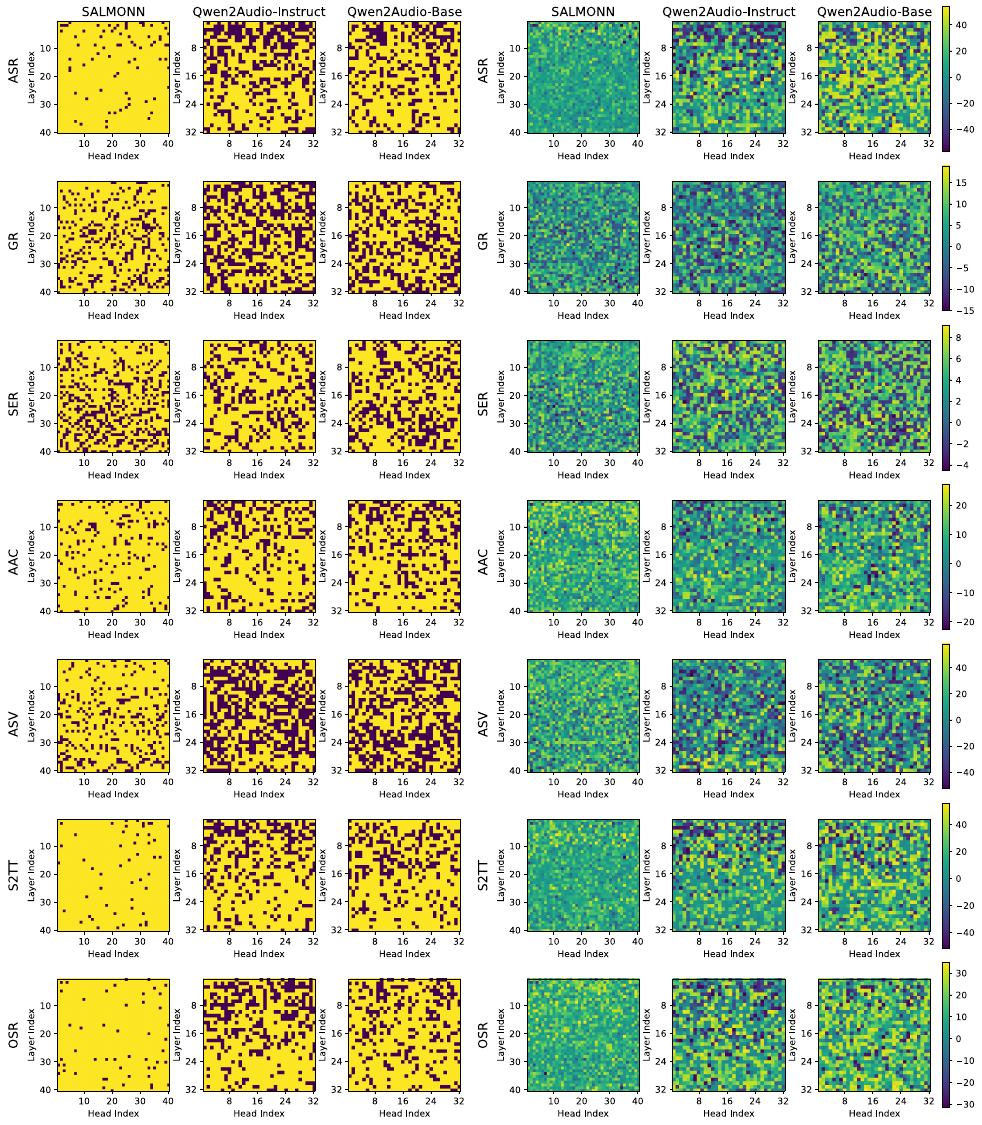}
    \caption{Visualization of masks for all models on all the 7 single tasks.
    Each row represents one task, and each column represents one LALM. 
    \textbf{Left}: discrete binary mask $\mathcal M$ in inference. The bright color denotes activated heads, while dark denotes masked heads.
    \textbf{Right}: the mask logits $\mathbf M$ that serves as the parameter for training.}
    \label{fig:all-masks}
\end{figure*}

Please refer to Fig. \ref{fig:all-masks} for visualizations of all the attention head masks corresponding to Table \ref{tab:single-salmonn}.

\begin{table*}[]
\centering
\small{
\begin{tabular}{@{}lccl@{}}
\toprule
\multicolumn{1}{c}{\textbf{Task}} & \multicolumn{1}{c}{\textbf{\begin{tabular}[c]{@{}c@{}}Training Set \\ \# Samples\end{tabular}}} & \multicolumn{1}{c}{\textbf{\begin{tabular}[c]{@{}c@{}}Test Set \\ \# Samples\end{tabular}}} & \multicolumn{1}{c}{\textbf{Example Output}} \\ \midrule
ASR (Automatic Speech Recognition) & 281,241 & 2,620 \& 2,939 & Peter had asked him of course for Matthew... \\
GR (Gender Recognition) & 28,539 & 2,620 & One of \{``Female'', ``Male''\} \\
SER (Speech Emotion Recognition) & 4,090 & 1,241 & One of \{``Neutral'', ``Happy'', ``Sad'', ``Angry''\} \\
ASV (Automatic Speaker Verification) & 523,411 & 37,611 & One of \{``Yes'', ``No''\} \\
AAC (Automatic Audio Captioning) & 48,073 & 964 & Motorcycle revving followed by a child crying. \\
S2TT (Speech to Text Translation) & 289,377 & 15,531 & \includegraphics[width=4cm]{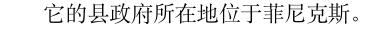} \\
OSR (Overlapped Speech Recognition) & 64,700 & 3,000 & Who had he but continued... Scientific breeding might... \\ \midrule
ASR \texttt{|} GR & 28,539 & 2,620 & Peter had asked him ... \texttt{|} Female \\
GR \texttt{|} ASR & 28,539 & 2,620 & Female \texttt{|} Peter had asked him ... \\
\{``ASR'': , ``GR''\} & 28,539 & 2,620 & \{``ASR'': ``Peter had asked...'', ``GR'': ``Female''\} \\ \bottomrule
\end{tabular}
}

\vspace{0.1in}

\small{
\begin{tabular}{@{}lccccccc@{}}
\toprule
\multirow{2}{*}{\textbf{LALM}} & \multicolumn{2}{c}{\textbf{Audio Encoder}} & \multicolumn{4}{c}{\textbf{LLM Backbone}} & \multicolumn{1}{c}{\multirow{2}{*}{\textbf{\begin{tabular}[c]{@{}c@{}}Total\\ \# Param.\end{tabular}}}} \\ \cmidrule(lr){2-3}\cmidrule(lr){4-7}
 & \textbf{Model} & \textbf{\# Param.} & \textbf{Model} & \textbf{\# Layers} & \textbf{\# Heads per Layer} & \textbf{\# Param.} & \multicolumn{1}{c}{} \\ \midrule
SALMONN & \makecell{Whisper-large-v2,\\BEATs-iter3+} & 0.775B & Vicuna-1.1 13B & 40 & 40 & 13.022B &  13.777B\\
\midrule
\begin{tabular}[c]{@{}l@{}}Qwen2Audio\\(Instruct \& Base)\end{tabular} & Whisper-large-v3 & 0.642B &  Qwen-7B &  32& 32 &  7.755B & 8.397B \\ \bottomrule
\end{tabular}
}
\caption{More dataset and model information for each task. There are two test sets for the ASR task: LibriSpeech~\cite{librispeech} test-clean \& test-other.
The encoder in Qwen2Audio is initialized on the Whisper-large-v3 model and further tuned, while SALMONN freezes the pretrained Whisper-large-v2 and BEATs-iter3+ models.}
\label{tab:dataset-more}
\end{table*}

\begin{table*}[htbp] 
\centering
\small{
\begin{tabularx}{\textwidth}{
  >{\RaggedRight\hsize=.5\hsize}X
  >{\centering\hsize=.5\hsize}X
  >{\RaggedRight\arraybackslash\hsize=2\hsize}X
}
\toprule
\multicolumn{1}{c}{\textbf{Model}} & \multicolumn{1}{c}{\textbf{Task}} & \multicolumn{1}{c}{\textbf{Instruction}} \\
\midrule
\multirow{14}{*}{SALMONN} & ASR & Recognize the speech, only output the transcription:  \\
 & GR & Recognize the speaker's gender, in one word: \\
 & SER & Describe the emotion of the speaker in one word. \\
 & ASV & Is only one person speaking in the audio?  \\
 & AAC & Please describe the audio. \\
 & S2TT & Translate the speech into Mandarin, only output the translated text:  \\
 & OSR & Please write down what you hear each person says. \\
 & GR\texttt{|}ASR & Given the speech utterance, first output the speaker's gender, then transcribe the spoken content. Separate the gender and transcription with `\texttt{|}', such as `Female\texttt{|}\texttt{<}actual content\texttt{>}'. \\
 & ASR\texttt{|}GR & Given the speech utterance, first transcribe the spoken content, then output the speaker's gender. Separate the transcription and gender with `\texttt{|}', such as `\texttt{<}actual content\texttt{>}\texttt{|}Female'. \\
 & \{``ASR'': , ``GR'': \}  & Given the speech utterance, output the transcription and speaker's gender, and format the results into json style. The keys for transcription and gender should be ``ASR'' and ``GR'' respectively. For example: \{``ASR'': ``actual content'', ``GR'': ``Female''\}. \\
\midrule 
\multirow{16}{*}{Qwen2Audio-Instruct} & ASR & Recognize the speech in English. Only output the transcription. Don't output other texts. \\
& GR & Recognize gender of the speaker in one word.\\
& SER & Recognize emotion of this utterance in one word: Happy, Sad, Angry or Neutral? \\
& ASV & Identify if there is only one person speaking in the audio clip. Only output Yes or No. \\
& AAC & Please describe the audio very briefly, using only short phrases. \\
& S2TT & Translate the English speech to Mandarin. Only output the translated text. \\
& OSR & Please write down what you hear each person says. Only output the transcription of each person in any order. \\
& GR\texttt{|}ASR & Given the speech utterance, first output the speaker's gender, then transcribe the spoken content. Separate the gender and transcription with `\texttt{|}', such as `Female\texttt{|}\texttt{<}actual content\texttt{>}'. Only output gender and content (connected with `\texttt{|}') and don't output other texts.\\
& ASR\texttt{|}GR & Given the speech utterance, first transcribe the spoken content, then output the speaker's gender. Separate the transcription and gender with `\texttt{|}', such as `\texttt{<}actual content\texttt{>}\texttt{|}Female'. Only output content and gender (connected with `\texttt{|}') and don't output other texts. \\
& \{``ASR'': , ``GR'': \} & Given the speech utterance, output the transcription and speaker's gender, and format the results into json style. The keys for transcription and gender should be ``ASR'' and ``GR'' respectively. For example: \{``ASR'': ``actual content'', ``GR'': ``Female''\}. \\
\midrule
\multirow{16}{*}{Qwen2Audio-Base} &ASR& Recognize the speech in English: \\
&GR& Recognize gender of the speaker in one word: \\
&SER& Describe the emotion of the speaker in one word: Happy, Sad, Angry or Neutral?\\
&ASV&  Identify if there is only one person speaking in the audio clip. Only output Yes or No:\\
&AAC&  Please describe the audio very briefly, using only short phrases:\\
&S2TT&  Translate the English speech into Mandarin: \\
& OSR &Please write down what you hear each person says. Only output the transcription of each person in any order:\\
& GR\texttt{|}ASR & Given the speech utterance, first output the speaker's gender, then transcribe the spoken content. Separate the gender and transcription with `\texttt{|}', such as `Female\texttt{|}\texttt{<}actual content\texttt{>}'. Only output gender and content (connected with `\texttt{|}') and don't output other texts:\\
& ASR\texttt{|}GR & Given the speech utterance, first transcribe the spoken content, then output the speaker's gender. Separate the transcription and gender with `\texttt{|}', such as `\texttt{<}actual content\texttt{>}\texttt{|}Female'. Only output content and gender (connected with `\texttt{|}') and don't output other texts: \\
& \{``ASR'': , ``GR'': \}& Given the speech utterance, output the transcription and speaker's gender, and format the results into json style. The keys for transcription and gender should be ``ASR'' and ``GR'' respectively. For example: \{``ASR'': ``actual content'', ``GR'': ``Female''\}: \\
\bottomrule
\end{tabularx}
}
\caption{Detailed instructions for each LALM and each task, for Table \ref{tab:single-salmonn} and \ref{tab:composite-salmonn}.}
\label{tab:instructions}
\end{table*}

\end{document}